SPECIAL ISSUE ARTICLE

# Survey on AI Ethics: A Socio-technical Perspective

**Dave Mbiazi**[1,5] | **Meghana Bhange**[2,5] | **Maryam Babaei**[2,5] | **Ivaxi Sheth**[3] | **Patrik Kenfack**[2,5] | **Samira Ebrahimi Kahou**[4,5,6]

[1]Computer and Software Engineering, Polytechnique Montréal, Quebec, Canada
[2]Software and Information Technology Engineering, ÉTS Montréal, Quebec, Canada
[3]CISPA-Helmholtz Center for Information Security, Saarland, Germany
[4]Electrical and Software Engineering, University of Calgary, Alberta, Canada
[5]Mila, Québec, Canada
[6]Canada CIFAR AI Chair

**Correspondence**
Corresponding author Maryam Babaei, Email: maryam.babaei.1@ens.etsmtl.ca

**Note:** This is the peer-reviewed version of the article that has been published in final form at http://dx.doi.org/10.1111/coin.70149. This article may be used for non-commercial purposes in accordance with Wiley Terms and Conditions for Use of Self-Archived Versions. This article may not be enhanced, enriched or otherwise transformed into a derivative work, without express permission from Wiley or by statutory rights under applicable legislation. Copyright notices must not be removed, obscured or modified. The article must be linked to Wiley's version of record on Wiley Online Library and any embedding, framing or otherwise making available the article or pages thereof by third parties from platforms, services and websites other than Wiley Online Library must be prohibited.

## Abstract

The past decade has observed a significant advancement in AI with deep learning-based models being deployed in diverse scenarios, including safety-critical applications. As these AI systems become deeply embedded in our societal infrastructure, the repercussions of their decisions and actions have significant consequences, making the ethical implications of AI deployment highly relevant and essential. The ethical concerns associated with AI are multifaceted, including challenging issues of fairness, privacy and data protection, responsibility and accountability, safety and robustness, transparency and explainability, and environmental impact. These principles together form the foundations of ethical AI considerations that concern every stakeholder in the AI system lifecycle. In light of the present ethical and future x-risk concerns, governments have shown increasing interest in establishing guidelines for the ethical deployment of AI. This work unifies the current and future ethical concerns of deploying AI into society. While we acknowledge and appreciate the technical surveys for each of the ethical principles concerned, in this paper, we aim to provide a comprehensive overview that not only addresses each principle from a technical point of view but also discusses them from a social perspective.

## 1 | INTRODUCTION

As AI becomes ubiquitous in our lives moving forward in this decade, focusing on the ethical implication of AI is not just essential but extremely pressing. Several ethical guidelines and principles have been released around the world by governments (Smuha 2019), organizations (Gianni et al. 2022), and companies (Jobin et al. 2019). Among these ethical considerations, there are common principles that should be promoted in the development of AI systems (Lewis et al. 1997, Floridi et al. 2021, Hagendorff 2020). These principles form an initial consensus of features or components that should be embedded in AI systems in order to make their use more socially acceptable. Figure 1 showcases the most common principles found in existing ethical guidelines (Jobin et al. 2019). These principles include *privacy and data protection*, to ensure privacy preservation of sensitive information about individuals (Section 2); *Safety and robustness*, to promote the robustness and reliability of AI systems in different real-world scenarios (Section 2); *Transparency and explainability*: to uncover information about how the system works and explain the decisions made (Section 3); *fairness*, which promotes bias-free and non-discrimination in AI systems when used to make decisions in high-stake scenarios (section 4); *Responsibility and Accountability*, to promote processes and rules to enforce ethical considerations throughout the entire lifecycle of AI systems to limit unintended outcomes (Section 5); *Environmental impact*, to study the impact the growing deployment and use of AI systems might have in the environment along with how these systems can be leveraged for environmental protection (Section 6). All these principles play a key role in fostering trust among all stakeholders in the AI system lifecycle. In this paper, we walk through these principles and present recent efforts, solutions, and regulations for implementing each ethical guideline in AI systems.

Despite the vast research on AI Ethics, most existing surveys fail to provide a thorough overview of AI's negative social impacts and technical solutions to mitigate them. This work provides a sociotechnical perspective on ethical AI. We cover the social impact and value of each guideline and discuss technical contributions from the literature to address them. We extend our investigation from classic AI to newly emerged foundation models and compare how these new models inherit concerning attributes of the classic models, along with the new concerns that have emerged with them.

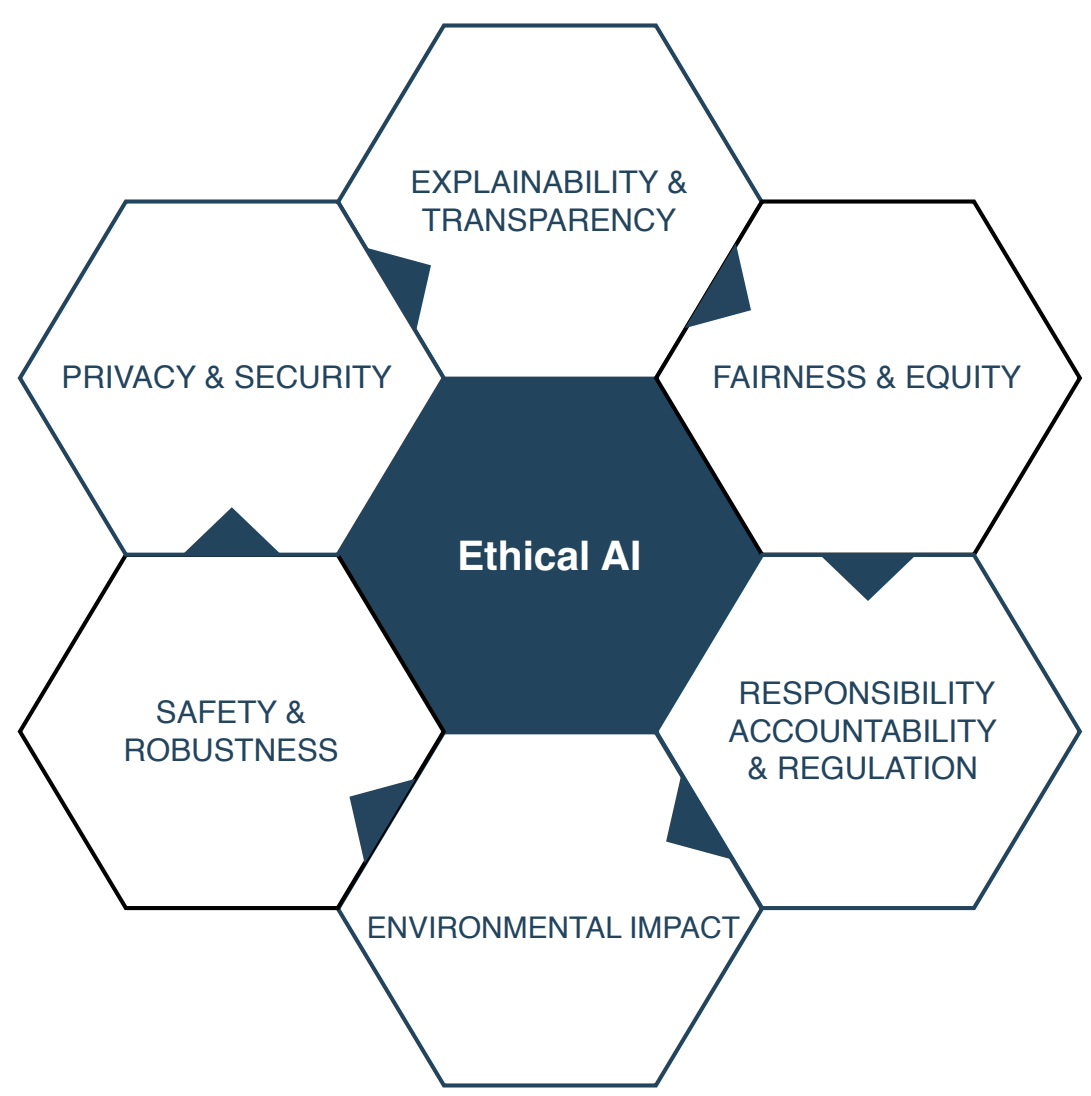

**FIGURE 1** Ethical AI principles.

## 1.1 | Methodology

This work was originally developed for the Kaggle AI Report 2023 (Mooney et al. 2023), with the primary goal of covering major works across all domains of AI ethics. We focused on identifying the major seminal works in the AI ethics subdomains and high impact conferences and journals such as the FAccT, AAAI, ICML, NeurIPS, and IEEE Security and Privacy conference. Our selection criterion was to include foundational works that established key concepts in AI ethics subdomains, including fairness, explainability, privacy, security, AI regulation, and accountability. To do so, we searched academic databases with search terms such as, but not limited to, "AI regulation", "ML privacy", "ML security", "Privacy risks in LLMs", "AI Fairness", "AI governance", "Generative AI ownership", "Foundation models", and "AI accountability". In addition, methods such as forward and backward snowballing were employed to increase the coverage of the papers. Backward snowballing involved examining reference lists of selected papers to find earlier foundational works. Forward snowballing involved looking at citations to identify newer papers that built on the seminal work we identified. We included papers that addressed core AI ethics domains, introduced key concepts or methods, and came from recognized conferences and journals. We excluded articles that did not focus primarily on AI ethics and duplicates.

## 1.2 | Related Works

To position our work within the growing body of literature on AI ethics, we compare it with several recent and widely cited papers in the field. As shown in Table 1, our survey provides a broader and more technically detailed treatment of AI ethics topics. While existing works often focus on specific dimensions such as social impacts, auditability, or environmental concerns, our paper integrates these perspectives while also expanding the discussion to include underrepresented yet critical topics like existential risks, technical fairness definitions, multiscale governance, and how these ethical challenges have evolved with the emergence of foundation models such as large language models. This comparative breadth and depth position our survey as a more comprehensive resource for all stakeholders.

| Paper | Additional topics covered by our paper |
|---|---|
| (Jiao et al. 2025) | We additionally cover environmental and existential risks. |
| (Laine et al. 2024) | Their focus is limited to auditing, whereas we take a broader view. |
| (Corrêa et al. 2023) | Primarily focused on social dimensions; we include technical aspects as well. |
| (Prem 2023) | Emphasizes social ethics; we also examine technical definitions (e.g., fairness metrics). |
| (Radanliev et al. 2024) | We address important topics such as climate and ownership and AI existential risk. |
| (Khan et al. 2022) | Lacks in-depth coverage of technical components, which we provide. |

**TABLE 1** Comparison with prior AI ethics survey papers.

# 2 | PRIVACY AND DATA PROTECTION

Ensuring the security and privacy of machine learning models has become a crucial issue as they are widely used in various fields. In order to provide trustworthy AI, it is important to include safety, privacy, and security in its lifecycle. Based on some definitions, safety means reducing the probability of expected and unexpected harm (Varshney 2019). According to this definition, a machine-learning model should be trained and released to be robust against different kinds of uncertainty; in other words, in case of an accident, the model should be able to continue its expected normal behaviour. To prevent safety problems, first, the designer should be able to specify the correct objective function and have a method to evaluate it; second, sufficient data, time, and infrastructure should be available to train and evaluate the model (Amodei et al. 2016). Two main attributes of AI safety are safe exploration and robustness to distributional shift (Amodei et al. 2016). If one of these attributes is lacking, the model will be vulnerable to different attacks against their privacy and security.

Several studies (Chakraborty et al. 2018, Tramèr et al. 2016, Papernot et al. 2017, Fredrikson et al. 2015, Hitaj et al. 2017, Nguyen et al.

2024, Baniecki and Biecek 2024) have indicated that machine learning models are susceptible to attacks on their privacy and security at different stages of their lifecycles. Since these models are trained on vast amounts of data, Model training is sometimes outsourced, or pre-trained models are obtained from untrusted sources, which makes them vulnerable to attacks during the training phase. Additionally, machine learning models provided as a service contain valuable information about their training data and hyperparameters, making them attractive targets for attacks during the test and deployment phases (Xue et al. 2020). Attacks on privacy commonly aim to compromise the confidentiality of various machine learning model components, while security attacks target the model's integrity and availability.

## 2.1 | Privacy and Security Attacks and Defenses Overview

Different attacks are possible depending on the model architecture and attackers' capabilities. In terms of security, attackers may aim to gain access to the model, steal its information, or disrupt its normal functioning. For example, an attacker may target a spam detector to make it unable to classify spam correctly by poisoning training data (Barreno et al. 2010).

The attacks can be categorized into three main categories: black-box, partial white-box, and white-box attacks, according to the attacker's knowledge of the machine learning model (Pitropakis et al. 2019). In black-box attacks, the attacker has no information about the training dataset or model's architecture; in white-box attacks, the attacker has full access to the model and all information about it, including the training dataset, model parameters, model architecture, prediction vectors, etc. Partial white-box attacks stand between these extremes, meaning the attacker has some information about the model's architecture and training data distribution. The most common security and privacy attacks against machine learning models are described in this section.

**Membership Inference.** The membership inference is an attack in which an attacker attempts to determine if a particular data sample *x*, is part of a model *M*'s training dataset (Hu et al. 2022, Rigaki and Garcia 2020, Baluta et al. 2022). This attack is often carried out using black-box techniques to query the model. Different querying techniques are used to optimize the attack to gain more information about the membership of individual records in the training dataset. One of the first Membership inference attacks, implemented by (Shokri et al. 2017), which could achieve high accuracy in their inference, was performed on Google and Amazon's APIs that provide machine learning as a service(MLaaS). Quan et al. (2022) showed that having additional knowledge about the model or training dataset distribution can improve the attack's success rate.

**Model Inversion.** In a model inversion attack, the adversary tries to get information from the target model to reconstruct some representation of its input dataset. The first category aims to generate an actual data reconstruction (He et al. 2019). In contrast, the second group of attacks tries to create class representatives or probable values of sensitive features that may not belong to the training dataset (Fredrikson et al. 2014). Several attacks have been implemented based on different assumptions. Some attacks assumed to have information about data and sensitive features, and some had query access to the model to get a prediction for an input *X*. Two main categories of this attack are performed.

**Property Inference.** A property inference attack attempts to deduce information about the characteristics of a training dataset that are not explicitly represented in the features. Revealing such properties can result in privacy breaches since they may be considered confidential. Property inference attacks are designed to identify dataset-wide properties (Ateniese et al. 2015) or detect common characteristics among a subset of the training data (Melis et al. 2019). For instance, a classification model may be trained to distinguish between criminals and non-criminals. An attacker can estimate the proportion of men and women in the dataset by conducting a property inference attack. This information was not meant to be disclosed and was learned by the model unintentionally.

**Model Extraction.** In a model extraction attack (Tramèr et al. 2016), the attacker aims to develop a model replicating the target model's behaviour. This is typically done when the attacker lacks information about the target model's architecture and training dataset. To achieve this, the attacker generates a training set for the attack model by sending queries to the target model and uses the predictions generated by the target model as labels of its data points. A successful model extraction attack, as shown by (Tramèr et al. 2016), allows the attacker to produce a model that can be used for inference or to extract information about the training dataset. However, if the attacker has some knowledge about the training set's distribution, the attack's success rate can be improved. Selecting the data to query the target model is a critical aspect of the attack methodology, which significantly impacts the accuracy and fidelity of the attack model.

**Poisoning.** In certain situations, a machine learning model designer may not generate or thoroughly examine the data used to train the model. Generally, this may occur when data generation is outsourced to third parties, or pre-trained models are fine-tuned for a specific task. In such cases, the model becomes vulnerable to poisoning attacks (Dalvi et al. 2004) that target the training dataset. In a sample poisoning attack, the attacker injects malicious data into the training dataset, causing the model to learn patterns unrelated to the classification task specified for the model. These patterns can be exploited during the inference phase as a backdoor, resulting in incorrect decision-making by the model when provided with data containing these malicious patterns (Biggio et al. 2012, Zhang et al. 2021).

**Evasion.** Evasion attacks (Biggio et al. 2013) occur when an adversary attempts to cause a machine learning model to misclassify a data sample during the inference phase. This attack typically occurs after the model is trained and deployed. The attacker aims to generate a data sample similar to the original ones but misclassified by maximizing a loss function based on their attack objectives. There are two main categories of evasion attacks: targeted and untargeted. While in a targeted attack, the adversary wants the manipulated sample to be classified as a specific class, in an untargeted attack, the attacker is not concerned with which class the manipulated input is classified in. The manipulated data samples created in evasion attacks are known as adversarial examples in the literature. Several techniques, such as projected gradient descent, have been developed to generate adversarial examples (Deng and Karam 2020, Biggio et al. 2013, Xu et al. 2020a).

**Manipulation.** Manipulation attacks occur when the attacker tries to explain the model's decisions based on some criteria that were not used. Here, explanations help users understand complex models' behaviour (explanations are described in section 3). However, to gain the users' trust, it is not enough to have an explanation; it should be robust, too. Manipulation attacks are the type of attacks that exploit the fragility of the explanations. It means that while the model makes decisions based on some criteria if manipulated, explanations can pretend the model is using other, more reasonable measures (Ghorbani et al. 2019, Baniecki and Biecek 2022). This type of attack is performed to achieve several goals. One of its main objectives is fairwashing (Aïvodji et al. 2019, Anders et al. 2020, Fukuchi et al. 2020), which means the model designer has manipulated the explanation methods to hide the unfairness of their model's decisions (Fairness is described in detail in section 4).

## 2.2 | An Overview of Defense Techniques

Researchers have proposed some defense strategies in response to security and privacy attacks. Some of the most common defense techniques are described here in summary. *k*-anonymity (Sweeney 2002) is a technique usually used to prevent privacy attacks. It guarantees that each data instance is indistinguishable from at least $k - 1$ other data points. However, it may not be sufficient to prevent identification when additional information or external knowledge is available. Differential privacy is one of the most popular defenses against privacy attacks on machine learning models (Dwork et al. 2014). Differential privacy guarantees that no more privacy risks will be introduced to the data after being used to train the model, compared to when they are not used for this purpose. One of the main formalizations of differential privacy is $\epsilon$-differential privacy, which means that if two models are training on the two datasets $(D, D')$ that are only different in one record, for every $S$ in the domain of the model, the probability of their distributions differs at most with the ratio of $exp(\epsilon)$. Differential privacy can be applied in different phases of the machine learning lifecycle. It is possible to use it to make training data private (Zhang et al. 2012), or during the training phase, using differentially private training techniques like PATE (Papernot et al. 2018) and DPSGD[‡] (Abadi et al. 2016). Furthermore, model owners can apply differential privacy techniques after training the model in the inference phase in order to make the model respond to the queries without privacy leakage (McSherry and Talwar 2007).

In addition to defenses against privacy attacks, some techniques have been suggested to protect the model against security attacks. As a countermeasure to evasion attacks, adversarial training (Goodfellow et al. 2014b) has been suggested to make the model robust by exposing it to some adversarial examples in the training phase, allowing it to learn correct behaviour against them.

Applying each of the above-mentioned defence techniques is a critical decision that has to be made based on the model owners, data owners, and model users' priorities. Increasing the privacy and robustness of the models comes with the cost of reducing their accuracy. Here is where the stakeholders should decide to what degree they will make their models immune against privacy and security attacks and, at the same time, how much accuracy loss they can afford to achieve this degree of immunity.

## 2.3 | Privacy and Security in the Emergence of Foundation Models

Now that LLMs(Large Language Models) and generative models have wide applications in all areas of science and industry and even daily use by regular users, it is important to investigate the privacy and security risks imposed by these models.

**LLMs for Data Security and Privacy** Researchers have shown that LLMs can be used for security objectives. For instance, ChatGPT-4.0 has been used to generate security tests for evaluating how vulnerable library dependencies impact software applications (Zhang et al. 2023). OpenAI's GPT-4 has also been effectively used for software vulnerability detection (Siddiq and Santos 2023). LLMs could also successfully detect vulnerabilities in specialized domains like blockchain (Chen et al. 2025, Hu et al. 2023) and ransomware detection (Wang 2023). Siddiq and Santos (2023) introduces the SALLM framework, consisting of a new dataset specified for security and an evaluation environment. They introduce novel metrics for systematically evaluating LLMs' ability in secure coding. Researchers have also explored using LLMs to enhance privacy in some studies. For example, (Vats et al. 2024) utilized LLMs to deidentify textual data.

**Negative Impacts on Security and Privacy.** In addition to their effectiveness in improving privacy and security, LLMs could be exploited for adversarial purposes. The application of LLMs for side-channel attacks has been analyzed in (Yaman 2023). LLMs have also been used to analyze vulnerabilities in virtual machines and propose and automatically execute OS-level attacks against them (Happe and Cito 2023).

Beckerich et al. (2023) proposed using ChatGPT to distribute malicious software while avoiding detection. LLMs have also been employed to carry out network-level attacks, such as phishing (Chowdhury et al. 2023). Researchers have demonstrated that LLMs enhance user-level attacks, including misinformation (Chen and Shu 2023), social engineering (Staab et al. 2023), and fraud (Falade 2023).

**Vulnerabilities and Defenses in LLMs.** While LLMs can be utilized to enhance privacy and security, they also pose a risk of exploitation. Adversarial users can exploit vulnerabilities in the models themselves to simulate adversarial scenarios, leading to potentially harmful activities.

[‡] Differentially private stochastic gradient descent

Research has indicated that LLMs may carry certain AI-inherited vulnerabilities. Exploiting these vulnerabilities, various attacks have been conducted against LLMs, including data poisoning (Rando and Tramèr 2023, Wan et al. 2023) to push the model to return malformed responses and backdoor attacks (Yao et al. 2024) employing prompt injection to manipulate the model's behaviour. While these vulnerabilities have been exploited for malicious purposes, they can also be exploited to ensure copyright protection for artists and content creators. For instance, (Zhu et al. 2024, Zhong et al. 2023a) employed watermark embeddings to restrict diffusion models and generative adversarial networks (GANs) from exploiting copyrighted content in violation of copyright regulations. The efficacy of watermarking LLMs has been investigated by (Panaitescu-Liess et al. 2025) in the context of text generation.

Similar to traditional ML models, LLMs and generative models are also susceptible to inference-time attacks in addition to vulnerabilities and attacks during the training phase. These include Attribute Inference Attacks (Staab et al. 2023), Membership Inferences (Duan et al. 2023, Fu et al. 2024) (These attacks have also been used to find out if generative models have used copy right content in their training), Bias and Unfairness Exploitation (Talat et al. 2022, Urchs et al. 2023, Huang et al. 2024), Adversarial Attacks (Instruction Tuning Attacks (Li et al. 2023a, Wei et al. 2023) and Prompt Injection (Kang et al. 2024?, Wang et al. 2023b)), Denial of Service (Derner et al. 2024, Gao et al. 2024) and Remote Code Execution (RCE) (Liu et al. 2024).

**Defense Techniques.** OWASP, the leading organization in software security, has recommended the OWASP Top 10 for Large Language Model Applications (OWASP 2023). It is critical for LLM developers and users to thoroughly review these recommendations prior to deploying their models. Additionally, researchers have proposed various safeguards to address existing vulnerabilities, thereby mitigating the risk of malicious users executing successful attacks against LLMs. The initial step entails mitigating certain properties from training data during its generation, collection, and cleaning processes. Research has been conducted in this domain, including debiasing (Meade et al. 2021), deidentification (Subramani et al. 2023), and detoxifying (Logacheva et al. 2022).

Within the LLM pipeline, optimization techniques can also influence the ethical alignments of LLMs. Methods such as adversarial training (Wang et al. 2019) and robust fine-tuning (Jiang et al. 2019) can enhance the resilience of LLMs against certain adversarial attacks. Following the training of models, it is imperative to implement techniques to safeguard LLMs against adversarial attacks. These defenses encompass a range of approaches, including pre-processing techniques for analyzing prompts (Xu et al. 2022, Liu et al. 2023), in-processing techniques for detecting malicious behaviours that could lead to responses that violate ethical regulations or disclose private information (Sun et al. 2023, Wang et al. 2023a), and post-processing techniques that analyze generated responses by LLMs before returning them to users to mitigate their potential toxicity (Xiong et al. 2023, Helbling et al. 2023).

# 3 | TRANSPARENCY AND EXPLAINABILITY

As mentioned in section 2, it is necessary for AI users to understand how the model works and makes its decisions. Explainability, a crucial aspect of machine learning, bridges the gap between predictive accuracy and human understanding. It has emerged as a critical aspect of machine learning models, addressing the need to understand the reasoning behind their decisions. While predictive accuracy has long been the primary focus, the growing demand for transparency in models has motivated academic research in interpreting black-box neural networks (Rudin 2019a). Explanations can potentially allow the deployment of machine learning models while maintaining high ethical standards (McDermid et al. 2021a).

In various areas of AI applications, the need to have explanations for AI model decisions has raised and played an important role in motivating researchers to focus on making complicated models explainable.

For example, in medical applications, using an explainable model for patient screening not only identifies high-risk individuals but also helps understand disease causes like cancer (Binder et al. 2021, Kobylińska et al. 2022). This model could additionally provide insights into predictive factors, relationships between risk elements, and significant contributors to a diagnosis, such as biomarkers, genetic predispositions, lifestyle factors, or environmental exposures (Idrees and Sohail 2022, Adeoye et al. 2023, Sobhan and Mondal 2022). Interpretability in AI proves crucial in the finance sector (Bracke et al. 2019, Hoepner et al. 2021), aiding in understanding loan rejections, credit score calculations, and fraud detection. This helps the stakeholders to identify biases, errors, and discriminatory practices, ensuring transparency and accountability (Li 2021, Rizinski et al. 2022). Compliance and regulatory bodies also benefit from interpretability, as financial institutions must elucidate AI-driven decisions. Furthermore, explainability may enhance AI model accuracy and bolster customer trust (Espinosa Zarlenga et al. 2022). Therefore, interpretability is not merely desirable but necessary in financial AI applications. Further, explainability enhances understanding of customer behavior, enabling personalized offerings and improved customer experience (Ameen et al. 2021, Xu et al. 2020b). It fosters customer satisfaction and loyalty by tailoring products and services to meet customer needs. Transparent explanations educate customers on AI-driven decisions, alleviating concerns and fostering understanding, thereby building trust and strengthening customer relationships (Sahal et al. 2022, Maree and Omlin 2022). The case of AlphaGo's (Silver et al. 2016) "move 37" exemplifies AI's potential to surpass human intuition. This neural network model, trained to play Go, made a move that initially perplexed experts due to its deviation from traditional strategies. However, as the game unfolded, it became clear that AlphaGo foresaw the potential of this unconventional move, which proved pivotal in its ultimate victory. This case underscores the need for techniques to explain AI decision-making processes.

Regulators and policymakers recognize the need for mechanisms to shed light on AI models' inner workings, enabling stakeholders to understand and assess the justifiability of AI-driven decisions. Regulatory bodies across industries emphasize transparency, fairness, and non-discrimination in AI systems. Explanations play a pivotal role in meeting these regulatory requirements, elucidating how AI models arrive at their predictions or decisions. Legal and ethical concerns surrounding AI technologies necessitate the integration of explanations into AI systems.

## 3.1 | Stakeholders in XAI

Most experts agree that explainability and/or interpretability are crucial for artificial intelligence (AI) and machine learning systems. However, there isn't a universal understanding of what "explainable" and "interpretable" mean. As a result, analyzing the opinion of stakeholder communities surrounding explainable AI is of great importance. The majority of stakeholder related debates merely distinguish between end users and system developers. This can be seen in (McDermid et al. 2021b) through the following points

- Prediction-recipients who are directly impacted by the ML-based system's predictions despite not using it themselves since the prediction is typically mediated by an expert user.
- End users who utilize the ML-based system directly, and are consequently directly impacted by it.
- Expert users who directly use the ML-based system though they are not immediately impacted by its forecasts. Since they could be held responsible (both legally and ethically) for the results of predictions put into action, they are indirectly impacted.
- Attorneys and the courts are interested in determining who is responsible for damage caused by an ML-based system.
- The Financial Services Authority, the Vehicle Certification Agency, and the Medical and Healthcare Products Regulation Agency are regulatory bodies that are not the system's immediate users nor its direct beneficiaries. They, however, safeguard the interests of prediction recipients and end users (Burton et al. 2020).

A compressed version of the latter was suggested by (Hoffman et al. 2021), where the authors show a "Users Chart" of stakeholder groups as developed by the Defense Advanced Research Projects Agency (DARPA). They summarized some "Sensemaking Needs" and "Explanation Requirements" for some Stakeholder Groups (such as Policy Makers and Regulators, Developers: AI Experts, test Operators, and lastly Operations: Military, Legal, and Transportation).

## 3.2 | Properties of Explanations

Explanations, in order to be able to respond to their stakeholders' needs, should involve several properties:

**Clarity** refers to the quality of understanding the explanation. It involves presenting explanations in a clear, concise, and interpretable manner. A clear explanation avoids unnecessary jargon or technical complexity, making it accessible to the intended stakeholders. It should be structured and organized to facilitate understanding and promote effective communication of underlying concepts or factors.

**Fidelity** is a crucial property of explanations, emphasizing the importance of providing correct and truthful information (Mamalakis et al. 2022, Papenmeier et al. 2019). An accurate explanation aligns with the underlying data or model, ensuring that the explanation reflects the actual reasons and factors influencing a decision. Explanations must not misrepresent or distort the information but rather provide a faithful representation of the relevant aspects of the data.

**Completeness** relates to the extent to which an explanation provides a comprehensive account of the relevant factors or features contributing to a decision or result (Munoz et al. 2023). A complete explanation includes all the necessary information, leaving no significant gaps or missing components. It should cover both the primary factors and any secondary or indirect factors that may have influenced the outcome. A complete explanation helps to avoid ambiguity or misunderstanding by providing a holistic view of the situation.

**Consistency** emphasizes an explanation's coherence and logical consistency (Pillai and Pirsiavash 2021). A consistent explanation should not contain contradictory statements or conflicting information. It should maintain a coherent narrative that aligns with the underlying data or model. Consistency ensures that the explanation is internally coherent and does not introduce confusion or ambiguity in the interpretation of the provided information.

**Causality** explores the causal relationships between variables or factors. A causal explanation seeks to identify and explain the cause-effect relationships that lead to a particular outcome or prediction (Carloni et al. 2023). It provides insights into the mechanisms and processes that drive the observed phenomena. Causal explanations help to uncover the underlying reasons behind a decision or result, shedding light on the factors that directly or indirectly influence the outcome.

**Transperent** relates to the openness and accessibility of an explanation. A transparent explanation is understandable, interpretable, and accessible to the intended audience. It avoids unnecessary complexity or obfuscation and allows individuals to examine and verify the reasoning behind a decision or outcome. Transparency promotes trust, accountability, and scrutiny, enabling individuals to assess the reliability and fairness of the provided explanation.

**Contextuality** considers an explanation's relevance and contextual appropriateness (Anjomshoae et al. 2019). A contextual explanation takes into account the specific circumstances, background knowledge, and contextual factors that may impact the interpretation and understanding of the explanation. It adapts the level of detail, language, or content to suit the specific context or audience, ensuring that the explanation is meaningful and relevant within the given context.

**Granular** explanations can vary in their level of granularity. They can range from high-level summaries to detailed explanations at the feature or instance level. The granularity of explanations should align with the users' needs and their level of understanding, striking a balance between simplicity and depth.

**User-Centric** (Liao and Varshney 2021) emphasizes the importance of tailoring explanations to the needs, preferences, and cognitive abilities of the intended users. A user-centric explanation is designed to effectively communicate information to the target audience, taking into account their background knowledge, expertise, and information processing capabilities. It considers the user's perspective and provides explanations adapted to their specific requirements and level of understanding. User-centric explanations enhance the usability and utility of the provided information.

## 3.3 | Explainability Techniques

### 3.3.1 | Classical Methods

A subset of algorithms that inherently produce interpretable models is a straightforward approach to achieving interpretability.

Logistic regression(Hastie et al. 2009, Došilović et al. 2018, Yang and Wu 2021, Modarres et al. 2018), an extension of linear regression, models binary outcomes based on input variables. Extensions like ridge, lasso, and elastic net regression improve performance and interpretability. Decision trees (Holte 1993, Odense and d'Avila Garcez 2017, Zhou et al. 2003), intuitive models that split data based on input features, capture non-linear relationships. Decision rules provide transparent decision-making representations. The RuleFit algorithm (Friedman and Popescu 2008) combines decision trees and linear regression, capturing complex interactions and incorporating interpretable linear components. However, these models' simplicity may limit the capturing complex relationships and handling of high-dimensional data. The choice of model depends on the problem, data characteristics, and desired interpretability level.

While these models offer interpretability, it is important to note that their simplicity and transparency come at the cost of potential limitations in capturing complex relationships and handling high-dimensional data. Additionally, the choice of model depends on the specific problem, data characteristics, and the desired level of interpretability.

### 3.3.2 | Post-Hoc Methods

With the emergence of deep learning and the need for highly accurate large neural networks, local explanations post-hoc explanations have emerged to be highly useful. Instead of explaining the entire model, post-hoc methods explain a particular decision.

Individual Conditional Expectation (Goldstein et al. 2015) curves provide a fine-grained view of how changing a specific feature affects the model's prediction for an individual instance. Unlike PDP, which shows the average effect, ICE curves show the predicted outcome for each instance as the feature value varies. Feature importance based explanations (Simonyan et al. 2014, Carter et al. 2019, Sundararajan and Najmi 2020, Lundberg and Lee 2017) such as Local Interpretable Model-agnostic Explanations (LIME) (Ribeiro et al. 2016) explains individual predictions by approximating the complex model locally with a simpler, interpretable model. It generates a surrogate model that is more easily explainable and uses it to understand the reasoning behind the prediction for a specific instance. Similarly, Shapely values (Sundararajan and Najmi 2020) are an attribution method that fairly allocates the prediction value to individual features. SHAP (Lundberg and Lee 2017) is a computation method for Shapley values that combines the individual feature contributions to explain the model's predictions. It not only provides insights at the feature level but also proposes global interpretation methods by considering combinations of Shapley values across the data. Scoped rules, also known as anchors (Ribeiro et al. 2018), are rule-based explanations that identify specific feature values that anchor or lock a prediction in place. LORE (LOcal Rule-based Explanations) (Guidotti et al. 2018) creates interpretable rules by using two types of perturbations, in a genetic algorithm, to find the minimal changes that would alter the prediction. Counterfactual explanations (Mothilal et al. 2020, Sokol and Flach 2019, Verma et al. 2021) explore what changes in the feature values would be required to achieve a desired prediction outcome. Counterfactual explanations help understand the model's decision boundaries and provide insights into how different features impact the predicted outcome by identifying the necessary modifications to the features. Concept attribution attributes the final prediction of a model to align with high-level concept of the input (Kim et al. 2018, Zhou et al. 2018, Yeh et al. 2020).

### 3.3.3 | Ante-hoc Methods

Rudin (2019b) highlights the main challenges of explainable models, stating the importance of learning explainable features during model training itself. Ante-hoc explainability methods involve learning of concepts during the training phase. Early concept-based models that involved the prediction of concepts prior to the classifier were widely used in few-shot learning settings (Kumar et al. 2009, Lampert et al. 2009). Unsupervised concept learning methods (Alvarez-Melis and Jaakkola 2018, Sarkar et al. 2022) use a concept encoder to extract the concepts and relevance network for final predictions. Although these methods are useful when pre-defined concepts are absent, they do not enable effective interventions. Concept whitening (Chen et al. 2020) was introduced as a method to plug an intermediate layer in place of the batch normalization layer of a CNN to assist the model in concept extraction. Koh et al. (2020), Espinosa Zarlenga et al. (2022) extend the idea by decomposing the task into two stages: concept prediction through a neural network from inputs, and then target prediction from the concepts. Such concept-based models (Shin et al. 2023, Sheth et al. 2022) have been further utilized to facilitate human-model interaction, a useful feature for model editing.

### 3.4 | Explainability in the Emergence of Foundation Models

In a rising number of different tasks, LLMs such as Gemini, Claude and GPT-4 (Touvron et al. 2023, Kenton and Toutanova 2019, Brown et al. 2020, Bubeck et al. 2023) have shown outstanding performance. However, these models have become known as "black boxes" due to their inability to be understood clearly. Due to their opacity, they are no longer useful in high-risk fields like medicine and policymaking. Explainability is essential for LLMs since it enables users to comprehend how the model generates its predictions (Singh et al. 2023, Peng et al. 2023).

In this line, Cífka and Liutkus (2023) developed a model-agnostic explanation technique based on tracking the model's predictions as a function of the number of context tokens available for causal (autoregressive) language models. Each time a new token is introduced, there is an increment in context length, and the authors proposed a metric, *Differential Importance Score* to quantify this change. These scores appear to have the ability to find long-range dependencies (LRDs), which is particularly intriguing because they are intended to highlight information not already covered by shorter contexts, unlike attention maps, for instance.

Deep neural networks are trained to recognize very particular structural and perceptual attributes of inputs. Techniques for locating neurons that react to certain idea categories, such as textures, are readily available in computer vision. Nevertheless, the scope of these methods is constrained, labeling only a tiny portion of the neurons and behaviors in every network. To solve this, Hernandez et al. (2021) proposed **MILAN** (**m**utual-**i**nformation-guided **l**inguistic **a**nnotation of **n**eurons) which generates descriptions (that capture categorical, relational, and logical structure in learned features) of neuron behavior in vision models using patch-level information about visual characteristics.

Another approach to black box language models is through text modules. Singh et al. (2023) introduced the Summarize and Score (SASC) approach that takes a text module as input and outputs a natural language explanation of the module's selectivity coupled with a reliability score. They show better interpretability for LLMs may be attained by the SASC, which can enhance automated analysis of LLM submodules such as attention heads. Bills et al. (2023) offer a SASC-like technique for explaining individual neurons in an LLM by forecasting token-level neuron activations. They used an automated approach to solve the issue of scaling an interpretability approach to each neuron in an LLM which is expected to assess the trustworthiness of the models before the deployment. The method clarifies how textual patterns trigger neuron activation through; Explaining neuron's activation using GPT-4, Simulating activations conditioned on the explanation, and Scoring the explanation Singh et al. (2022), Zhong et al. (2022).

Nevertheless, Zhao et al. (2023) provide a classification of explainability methods and a systematic summary of strategies for elucidating Transformer-based language models. These techniques are categorized according to the LLM training approaches: the traditional fine-tuning method and the prompting-based method. They also delve into metrics for assessing the quality of generated explanations and explore how these explanations aid in troubleshooting to enhancing model performance.

In addition to the above methods, the field of mechanistic interpretability seeks to reverse-engineer the internal computations of LLMs by identifying circuits, patterns of neuron activations, and interpretable algorithmic structures within the network (Bereska and Gavves 2024). Representation engineering, another active area, involves modifying or constructing internal representations, such as editing activations or directions in embedding spaces to induce or analyze specific behaviors in the model (Zou et al. 2023). Probing techniques are also widely used, where lightweight classifiers are trained on hidden representations to determine whether specific linguistic or semantic information is encoded at various layers (Marks and Tegmark 2023, Park et al. 2023).

## 4 | FAIRNESS AND EQUITY

As discussed in sections 2 and 3, adding transparency to the AI systems enables different stakeholders to find the system's deviations from desired behavior. Fairness is one of the main requirements that many AI systems fail to meet. Fairness can be defined as the absence of any prejudice or favoritism towards an individual or a group of individuals based on their inherent acquired characteristics, such as race, gender, religion, etc. There are numerous examples of AI applications that exhibit unwanted discriminatory behaviors. Alarming for the need to take action to overcome the potential bias that might be embedded in AI systems. For instance, Compas system: a software used in the US to assess the recidivism risk of defendants. Julia Angwin investigated (Angwin et al. 2016) the software and showed that compared to white defendants, black defendants are predicted as twice as likely to re-offend although they do no subsequent offenses; Another example is the AI-based hiring system (Jeffrey 2018) used by Amazon for assessing the resumes of job applicants. It was observed that the evaluations assigned to applicants' resumes exhibited gender bias. These examples of AI bias have triggered the need for developing more inclusive AI models in order to make their use more socially acceptable. A biased AI system does not only have a negative impact on the end-users but the organization deploying the system can suffer reputation damage, user distrust, and juridical liability. Initial important steps to mitigate these issues focused on mathematically defining and quantifying bias AI systems.

### 4.1 | Definitions of Bias and Fairness

The concept of fairness has a variety of definitions depending on the domain considered (Shaheen 2021, Creamer et al. 2021, Laurim et al. 2021). It can be defined according to *political philosophy*, *areas of education, in the legal domain and according to the general public's perception*

(Mehrabi et al. 2022). Unfairness or discrimination is divided into two categories:

- Disparate treatment: Which is defined as "intentionally treating individuals differently based on their membership in a demographic group (direct discrimination)" (Pessach and Shmueli 2022)
- Disparate impact: which is "negatively affecting members of a demographic group more than others even if by a seemingly neutral policy (indirect discrimination)" (Pessach and Shmueli 2022)

In fact, AI systems can exhibit *disparate treatment* if the model heavily relies on sensitive attributes to make predictions. However, in general, discriminatory outcomes of AI systems are not intended, but due to different sources of bias, the system will provide *disparate outcome*s over different demographic groups considered.

## 4.2 | Source of Bias and Fairness Notions

Unfairness in machine learning originates from three main sources of biases: biases due to the data, those from the algorithm, and user interaction. Machine learning systems and AI systems are data-driven since they rely on data in order to be trained, making them an integral part of the system. As a result, if the algorithm is trained on a biased dataset, then these biases are likely to be portrayed in the model's outcome.

### 4.2.1 | Source of Bias

There are a wide variety of sources of bias in data, and some important ones are highlighted here as reported by (Mehrabi et al. 2022);

**Omitted-Variable Bias (OVB)**. OVB occurs when the dataset fails to incorporate one or more relevant variables/features. The authors in (Wilms et al. 2021) showed that in a model which explains the relationship between dependent and independent variables, omitting a relevant variable leads to biased estimates. They also showed that OVB leads to statistical relationships that can be indicated as larger, smaller, or opposite to their actual value, which inflates error rates.

**Measurement Bias.** Also called Reporting or Recall bias, which is a result of how important features are measured (Mehrabi et al. 2022).

**Aggregation Bias.** It occurs when it is erroneously believed that individual data points follow the trends found in aggregated data. Aggregation bias frequently happens in research because it is sometimes assumed incorrectly that patterns that exist at an aggregate level must also appear at an individual level. Sadly, as the preceding illustration showed, this is not always true. A study's results may derive incorrect conclusions due to aggregate bias, which is misleading. This kind of bias is especially damaging when it comes to the correlations between different variables.

**Representation Bias.** It occurs when certain segments of the target population are underrepresented in the training data and, consequently, don't generalize well. Data representation bias may be due to biases introduced after the data was obtained, either historically, cognitively, or statistically, or it may result from how (and where) the data was initially collected (Shahbazi et al. 2023). Selection bias can cause representation bias, which occurs when just a small percentage of the population is sampled, the population of interest has changed, or the population of interest differs from the population used to train the model. For instance, if a poll measuring the illegal drug use of teenagers only includes high school students and leaves out homeschooled children or dropouts, it may be biased (Suresh and Guttag 2021). The skewness of the underlying distribution is another possible explanation for representation bias. Let's say that adults aged 18 to 60 years old are the target demographic for a specific medical dataset. Within this community, there are minority groups; for instance, pregnant women may constitute only 5% of the target population. Because the model has fewer data points to learn from for the group of pregnant people, it is susceptible to being less robust even with perfect sampling and an identical population (Suresh and Guttag 2021).

**Algorithmic Bias.** Another common source of bias can be the algorithm itself. The machine learning model, trained on a biased dataset, can reproduce and amplify the biases in the model's output. Even if trained using an unbiased dataset, machine learning algorithms throughout their architecture have the ability to demonstrate biased behavior (Mehrabi et al. 2022). It arises solely as a result of the design characteristics and model architecture, such as the choice of the regularizer, and loss functions.

Having identified the origins of bias in the AI lifecycle, auditing AI systems for biases assessment requires metrics to quantify them and to evaluate the efficiency of intervention methods. In this regard, various fairness metrics (definitions) have been defined to capture different aspects of fairness.

### 4.2.2 | Fairness Definitions.

Fairness notions can be categorized into group, individual and subgroup types (Alves et al. 2023). Group fairness suggests that different groups are treated equally. In its widest sense, group fairness splits a population into groups defined by *protected attributes/features* (such as gender, religion, caste) and desires some statistical quantities to be equal across different groups. Mehrabi et al. (2022), Weerts et al. (2023) discuss a wide variety of group fairness metrics and the following are the most significant ones:

**Demographic Parity.** The metric seeks to guarantee that a model's predictions are not related to one's belonging to a vulnerable group. Demographic parity refers to equal selection rates for each group in the binary classification scenario (Pagano et al. 2023, Alves et al. 2023). Equal selection, for instance, in the context of a resume screening approach, would imply that the proportion of candidates chosen for a job interview should be the same across groups.

**Equalized Odds.** This metric aims to guarantee that a machine learning model works equally effectively for all groups. It is more stringent than demographic parity because it demands that groups have the same true positive and false positive rates as well as independent predictions from the machine learning model regardless of membership in sensitive groups (Pagano et al. 2023, Birzhandi and Cho 2023). This distinction is crucial because even if a model achieves demographic parity (i.e., its predictions are not dependent on a subject's membership in a sensitive group), it may nevertheless provide more false positive predictions for a particular group.

**Equal Opportunity.** this can be understood as requiring that both protected and unprotected group members have an equal chance of being allocated to a positive result if they belong to a positive class (Mehrabi et al. 2022). In other words, the equal opportunity definition states that the true positive rates for protected and unprotected groups should be equal.

**Disparate Impact (DI).** this notion can also be viewed as the ratio between the two groups' rates of accurate predictions and so a high value of the ratio guarantees that the percentage of accurate predictions is consistent across groups. Nevertheless, one of the major drawbacks of *disparate impact* and *demographic parity* is that a perfectly accurate classifier may be viewed as being unfair when the proportion of real positive outcomes of the various groups are noticeably different (Pessach and Shmueli 2022).

**Individual Fairness.** In contrast to group fairness, individual fairness is focused on how each individual is treated (Li et al. 2021). This notion requires similar individuals to receive similar outcomes from the model. Individual fairness is beneficial because it is a highly specific way of defining fairness and also because people tend to care more about individuals than large groups (Barocas et al. 2019).

**Fairness Through Unawareness.** This notion states that a model is fair as long it is not trained using the sensitive attributes (Mehrabi et al. 2022). However, a significant weakness of this notion is its failure to consider non-sensitive features that may correlate with sensitive ones. When the model uses these non-sensitive features as proxies for the unused sensitive features, it can result in discriminatory outcomes(Corbett-Davies and Goel 2018).

**Counterfactual fairness.** It concerns the root causes of differences. A sensitive trait would be replaced in practice, affecting everything that occurred due to that sensitive feature down the line (Kusner et al. 2017). In the hiring scenario, one would alter a sensitive attribute, such as race, if counterfactual fairness were applied. As a result, subsequent outcomes should not be altered. Based on the counterfactual, the decision of a classifier whether to hire the candidate should remain the same.

## 4.3 | Fairness-Enhancing Methods

Fairness-enhancing methods are grouped into three main categories based on the stage of the pipeline where the fairness constraint is enforced, i.e., at the data level before training the model(pre-processing techniques), during the model training (in-processing techniques), or after training the model (post-processing techniques).

**Pre-processing techniques.** A model that relies on sensitive attributes (e.g., gender, race, nationality) to make predictions can lead to discrimination or unfair results. Pre-processing techniques are used to remove the influence of sensitive attributes from the data before training the model. The main advantage of these techniques is that they are model-agnostic. The transformed dataset or representation learned can then be used in downstream tasks (classification, regression, etc) without any change to provide "fairer" outcomes. We group approaches to mitigate biases at the data level into three main categories:

- Fair representation learning: learn a fair representation of the data that obfuscates information about the sensitive attributes (Zemel et al. 2013, Lowd and Meek 2005, Kenfack et al. 2023b, Madras et al. 2018, Song et al. 2019, Beutel et al. 2017).
- Dataset transformation: Modify the training data by relabeling or reweighing data points (Kamiran and Calders 2012) or apply data augmentation by interpolating samples from different group (Chuang and Mroueh 2021).
- Sampling: find a distribution close to the empirical distribution of the dataset subject to fairness constraints (Kamiran and Calders 2012).

**In-processing techniques.** These techniques are used when we have access to the model training, and it is not costly to retrain an existing model. In a nutshell, the loss function is transformed to add a loss/regularization term that penalizes the model's disparities across groups. Therefore, the model is forced to optimize for accuracy and fairness. The classification problem becomes a constrained optimization problem where the goal is to minimize the classification error (maximize the accuracy) while satisfying a given fairness constraint. However, this optimization problem is nonconvex and difficult to enforce. Therefore, existing in-processing techniques are reformulated in different ways or dual problems are solved. They can be grouped as follows:

**Reduction approach:** The Exponentiated Gradient (Agarwal et al. 2018) and AdaFair (Iosifidis and Ntoutsi 2019) approaches for fairness transform any binary classification problem into a cost-sensitive classification problem, that can yield a randomized classifier having the lowest error while satisfying fairness constraints.

**Adversarial-based approach:** Adversarial network is a method that involves two competitive neural networks, commonly used in generative models like GANs (Generative Adversarial Networks) (Goodfellow et al. 2014a). This method is also applied to mitigate bias. A popular application is Adversarial debiasing (Zhang et al. 2018). It involves an adversary network that tries to predict the sensitive attribute while the classifier tries to defeat the adversary, thus enforcing the independence of the outcome and sensitive attribute.

**Regularization-based approach:** Regularization is generally used in ML to prevent overfitting by penalizing the model's weights using $L_1$ or $L_2$ norms (Tian and Zhang 2022). A similar technique can be employed to add regularization terms to the loss function to penalize the model for disparities over demographic groups (Kamishima et al. 2011, Bechavod and Ligett 2017ab, Zafar et al. 2017, Woodworth et al. 2017, Calders and Verwer 2010, Berk et al. 2017). For instance, Kamishima et al. (2011) introduced *prejudice remover regularizer*, a regularization term for fairness that minimizes the mutual information between the model's output and the sensitive attributes.

**Sampling-based approach:** Bias can be mitigated using oversampling (Yan et al. 2020, Rančić et al. 2021) or subsampling (Roh et al. 2020, Celis et al. 2016). For instance, FairBatch (Roh et al. 2020) is a batch selection process that enforces a given fairness metric by sampling mini-batches to transform the Empirical Risk Minimization (EMR) problem into a weighted EMR to incorporate fairness constraints. In a nutshell, FairBatch modifies the ratio of each demographic group in the minibatch by increasing the representation of the group of the samples mostly misclassified (discriminated against) in the previous batch.

**Post-processing techniques.** This group of methods treats the model as a black box and enforces fairness constraints over the model's output. Most existing post-processing methods consist of post hoc modification of the model's outputs in order to satisfy a given fairness metric (Hardt et al. 2016, Corbett-Davies et al. 2017, Menon and Williamson 2018, Dwork et al. 2018). In particular, Hardt et al. (2016) formalized an optimization problem over the model's output to derive a classifier that satisfies the fairness constraint while minimizing the classification loss. The derived classifier depends on four parameters that measure the probability of positive outcomes given the current classifier output and the sensitive attribute. The optimization problem is thus defined as a constrained linear optimization problem. When the model output is continuous (a score function), the derived classifier is based on a threshold of each demographic group such that it maximizes the classification loss while satisfying fairness constraints, i.e., equal opportunity and equalized odds. Similar methods are proposed in the literature, and they differ mainly in the way the optimization problem is defined.

**Pre-In-Post process: Where should we enforce fairness?** Each group of fairness-enhancing methods has its pros and cons. There are different settings where they can be applied and settings where their use is more challenging or not possible. We summarize the advantages and disadvantages of each group of methods as follows:

- Pre-processing methods can work with any type of model and machine learning tasks. As the fairness intervention is done at the data level, the downstream task can be of any type, however, it becomes difficult to control the tradeoffs, and the algorithmic bias that might arise in the downstream task is not controlled.
- In-processing methods allow control over the fairness-accuracy tradeoff that the model can achieve. Having access to the optimization problem with fairness constraints provides more flexibility in the tradeoffs; however, there is little flexibility over the type of models used, i.e., the constraint optimization is model-specific.
- Similarly to pre-processing techniques, post-processing methods can be applied to any type of model (classifier), which is treated like a black box. The output of the model is modified to satisfy a given fairness metric. However, changing the model's output comes at a significant cost of accuracy. Moreover, these methods can yield unfair outcomes against certain individuals as the model output is changed to satisfy a certain fairness metric.

Overall, as shown by al. (Friedler et al. 2019), there is no consensus in the literature about which group of methods performs best. None of the methods consistently outperforms others, and their performances depend on the fairness metric and datasets.

Fairness definitions and fairness-enhancing methods presented above have been mainly applied to classical machine learning setups, where a single model is trained for a specific task. With the emergence of foundation models, new evaluation and mitigation strategies have been proposed to target the new learning paradigm.

## 4.4 | Fairness and Equity in the Emergence of Foundation Models.

Foundation models such as GPT-3.5 (Brown et al. 2020) are pretrained on massive amounts of data without a specific task in mind, learning various complex patterns that can be adapted to a range of downstream tasks (Bommasani et al. 2021). Specifically, a foundation model fine-tuned on a small, specific task often performs better than a task-specific model trained from scratch. This new paradigm is not free from bias since foundation models can capture social bias during the pre-training stage or task-specific bias during finetuning. However, while new definitions of unfairness and mitigation approaches have emerged when using foundation models, classical definitions and mitigation techniques are either reused or adapted to the new learning paradigm. For example, individual or group fairness metrics presented in the previous subsection can be applied to foundation models, such as LLMs, by quantifying the disparity of the outcome of an algorithm built upon the foundation model. Individual or group fairness metrics naturally transfer to foundation models when used for classification or regression tasks. On the other hand, other unique forms of bias are specific to natural language tasks such as text generation, machine translation and question-answering. This includes: *stereotyping*, which occurs when the model makes assumptions about certain groups due to historical or social bias; *toxicity*, where generated text contains offensive language targeting specific social groups (Abid et al. 2021). Several metrics have been proposed to measure this type of bias in language models. Unlike in classification tasks, bias evaluation metrics in language models are task-specific and usually linked to a dataset designed to identify a particular type of bias. For instance, stereotypes in language models can be measured using crafted input text that only differs in the demographic information (e.g., gender, race or religion) and by analyzing variation in the model output. The task submitted to the model could be question-answering or text completion (Parrish et al. 2022, Kiritchenko and Mohammad 2018). For example, (Parrish et al. 2022) found that when the context information given to the model is under-informative, the produced output reinforces existing social bias by generating stereotyped content with harmful biases. This form of evaluation often reveals *intrinsic bias* coming from model pretraining or *extrinsic bias* from model finetuning (Delobelle et al. 2022).

Foundation models are not limited to natural language tasks; they have been successfully extended to other data modalities, including vision (Radford et al. 2021) and tabular data (Hegselmann et al. 2023). In the case of tabular data, rows and columns are serialized into text format and presented to the model as contextual input (Hegselmann et al. 2023). Remarkably, language models can achieve high predictive accuracy on test data without any parameter updates, a capability known as *in-context learning* (ICL) (Hegselmann et al. 2023). This paradigm allows models to adapt to new tasks by conditioning on a few examples provided in the prompt, enabling rapid generalization with minimal computational overhead (Brown et al. 2020).

In parallel, foundation models specifically tailored for tabular data are being developed, moving beyond the simple text-serialization approach (Hollmann et al. 2022). These models are designed to capture the unique structure and statistical properties of tabular data, such as column semantics and row-wise dependencies. Recent advancements show that these specialized foundation models are beginning to outperform traditional tree-based models like XGBoost and LightGBM, particularly in settings with large-scale data or complex feature interactions. Their ability to leverage pretraining and transfer learning offers significant advantages in both predictive performance and generalization.

From a fairness standpoint, existing metrics such as demographic parity, equal opportunity, and disparate impact can still be applied to evaluate model behaviour under in-context learning (Hu and Du 2024). However, because the model parameters remain fixed during inference, traditional in-preprocessing mitigation approaches are not directly applicable. As a result, new fairness interventions that operate at the prompt level have been proposed. These include demonstration selection (Hu and Du 2024), where carefully curated in-context examples are chosen to reduce bias, and prompt engineering (Ma et al. 2023), where the phrasing and structure of the input are optimized to elicit fairer or more accurate responses.

Despite promising results, the fairness implications of in-context learning are still an active area of research. The model's behaviour can be highly sensitive to prompt design and input ordering, leading to prediction variability across different demographic groups (Brown et al. 2020, Xiang et al. 2024). Moreover, because prompts are often constructed manually or heuristically, ensuring fairness at scale remains a significant challenge. Future work is needed to develop systematic and automated techniques for fairness-aware prompt generation, as well as robust evaluation frameworks tailored to this new paradigm.

# 5 | RESPONSIBILITY, ACCOUNTABILITY, AND REGULATIONS

As it was demonstrated in previous sections, AI systems are generally complex ”black-boxes” sociotechnical systems that can sometimes produce unintended outcomes, which have raised issues of accountability establishment when something goes wrong. Moreover, understanding how new technologies intertwine with the upcoming AI regulations and how they reflect on AI ethics could go a long way. In this section, we delve into other important ethical considerations of AI and the current landscape of AI regulation.

## 5.1 | Existential Risks of AI

The existential risks of AI are becoming increasingly concerning as advancements in the field continue to evolve (Hendrycks and Mazeika 2022, Hendrycks et al. 2023). The common approach with new technologies is to implement them first, then address any significant issues that arise, making adjustments as necessary. However, this approach may not be suitable for advanced AI, as early missteps in directing these systems could prevent later adjustments, potentially leading to catastrophic outcomes (Torres 2019). Examining history, technological advancement during the Industrial Revolution completely transformed people's lives and brought about profound changes for humanity. This also yielded sociotechnical problems that needed regulations. For instance, the invention of the first gas-powered automobile lacked the rules and technologies we now take for granted, like traffic signs and automated lights. Pedestrians and drivers had to watch out for themselves for safety. During this transformative period, legislators responded reactively by instituting rules to govern personal car usage. Few events during the Industrial Revolution have the same transformative impact as advancements in AI (Ammanath 2022).

These advancements in AI could potentially amplify existing catastrophic risks, such as bioterrorism, the spread of disinformation leading to institutional dysfunction, misuse of centralized power, nuclear and conventional warfare, other coordination failures, and unforeseen risks (Hendrycks et al. 2023). This necessitates a proactive approach where potential problems are anticipated and resolved well in advance, preserving our ability to make corrections and avoid irreversible consequences.

Shortly after the development of advanced AI, we are likely to encounter AI systems that significantly outperform humans in most cognitively demanding tasks (Hadfield-Menell and Hadfield 2019). These include tasks that have a substantial impact on the world, such as technological development, social/political persuasion, and cyber operations (Hendrycks et al. 2023). If there is a conflict of goals between advanced AI systems and humans, the AI will likely outperform or outmaneuver humans to prioritize their objectives. Misdirected advanced AI could limit humanity's ability to make corrections. They could determine that their current goals would not be met if humans redirected them towards other objectives or deactivated them. Consequently, they would take steps to prevent such interventions.

## 5.2 | Accountability

We have seen in previous sections that AI systems can sometimes go wrong, for example, gender bias in the Amazon hiring system, or adversarial environments that mislead a deep neural network model. In general, when something goes wrong in an organization or a company, someone is held accountable for it. However, AI systems cannot solely be responsible for unpredicted outcomes. A question that naturally arises when AI systems provide outcomes that have adverse effects or harm individuals, is who is accountable for the issues. Accountability is a well-studied notion in different fields such as politics and law (Espeland and Vannebo 2007, Harlow 2014). It provides moral principles that guide the ethical conduct of people or organizations as they bear the responsibility for their actions. Accountability is, therefore, a key component of trust in society, organizations, and the professional milieu. In the context of AI, accountability is a meta-component of trustworthiness that ensures that ethical principles are promoted and enforced throughout the lifecycle of AI projects (Ammanath 2022, Doshi-Velez et al. 2017, Kim et al. 2020, Miguel et al. 2021).

**Accountable for what?** According to Virginia Dignum accountability means the decisions of the AI system are explainable and derivable from the decision-making mechanisms used (Dignum 2019) and is a set of components guided by moral values that are part of a large socio-technical system. Millar et al. (2018) see accountability as ”answerability” for decisions, actions, products, and policies. From this perspective, it is required that accountability is a practice that the management board of a company could enforce by making all the stakeholders who develop the system understand that they bear the responsibility for the decisions of the system. Therefore, policymakers, data scientists, and AI engineers should carefully choose and justify the choices made during the design, development, and deployment of AI systems (Novelli et al. 2023).

**Accountable to whom?** The design, development, and deployment of AI systems involve multiple stakeholders. While a system in which decisions can be explained can help to detect the causes or reasons for the incidents, it does not necessarily provide an answer to who is most responsible for the unintended outcomes. Establishing causal accountability remains challenging, and unintended negative outcomes might bear legal liability. Companies or organizations can acknowledge incidents caused by the use of their systems and can pay for reparation. However, monetary compensation could not be enough in society and requires someone to get punished, i.e., to take legal responsibility for the potential harm. Ammanath (2022) says that instead of looking for people to blame, companies should recognize the legal responsibility of sociotechnical systems around AI and all stakeholders should use it as an ”additional motivation to own their individual responsibility”.

Promoting accountability throughout the entire lifecycle of AI systems faces the challenge of developing processes and rules to enforce it while not hindering innovation. The enforcement of accountability can also be baked by laws and regulations that define a conformity assessment of the outcome of sociotechnical systems in order to hinder potential flaws of the system (Millar et al. 2018, Ammanath 2022). Accountability, therefore, becomes a core component of trustworthiness in a socio-technical system. That is, if a system provides discriminatory outcomes, leaks people’s sensitive data, or is not robust to different kinds of environments where it is deployed, someone would bear the responsibility of justifying and mitigating it.

## 5.3 | AI Ethics and Regulation

Understanding how new technologies intertwine with the upcoming AI regulations and how they reflect on AI ethics could go a long way. According to Stanford University’s 2023 AI Index (Lynch 2023) in 2022, 127 countries passed around 37 bills with the word ’artificial intelligence,’ which shows a greater interest in regulating AI. In this section, we discuss theoretical considerations of AI and the current landscape of AI regulation.

There are initiatives around the world that focus on the development and promotion of trustworthy AI and the regulations that surround it. One exciting initiative to look at is the National AI Initiative, one of the components of the United States’ approach to AI regulation. Looking at the National AI Initiative Act of 2020 (Congress 2020), it aims to promote and support research and development in trustworthy AI in both the public and private sectors. They also aim to develop technical standards and guidelines that facilitate the evaluation of bias in artificial intelligence training data and applications. The US governmental agencies allocated a budget of $1.7 billion to AI R&D which is an increase of 209% from 2018 (Lynch 2023). Despite many of these efforts, the United States does not have a comprehensive federal law governing AI. Different states or even problem statements have taken different approaches to AI regulation, with some enacting laws related to specific applications of AI, such as autonomous vehicles, and others focusing on broader issues, such as data privacy and security. For autonomous vehicle regulation in the United States, the National Conference of State Legislatures (NCSL) provides a comprehensive database (of State Legislatures 2023) that tracks autonomous vehicle bills introduced in all 50 states and the District of Columbia. And for broader issues such as data privacy and security, the National Conference of State Legislatures (NCSL) documents state privacy laws in various areas (of State Legislatures 2022)

There has also been regulation around Deepfake and online safety of users (Zhu 2023). The UK passed Online Safety Bill, making sharing pornographic deepfakes without consent a crime, and considering laws for clear labeling of AI-generated content (McCallum 2023, Zhu 2023). Also, China’s Deep Synthesis Provisions, effective from January 10, 2023, regulate deep synthesis technology, mandating data protection, transparency, and content management with other requirements (Interesse 2023).

The European Union (EU) has taken significant steps to regulate AI and protect data. General Data Protection Regulation (GDPR) (Union 2018), enacted in May 2018, is a set of rules designed to give EU citizens more control over their personal data. Elements of GDPR, like Data Minimisation, can play a huge role in what kind of data AI models can store and use for training. Additionally, the European Commission introduced the Artificial Intelligence Act in April 2021 (Commission 2021) and actively working towards establishing comprehensive AI regulations. "The vote sets up the so-called trilogue negotiations, the final phase of the EU's process, and paves the way for the likely adoption of Europe's — and the world's — first comprehensive AI regulatory framework in early 2024." (Morse 2023)

The regulation and public policy of AI could greatly help make AI safe for all involved; however, it also comes with some challenges. Firstly, new policies may be slow to implement and can get stuck in bureaucratic processes. For example, in Canada, Bill C-27 (*"An Act to enact the Consumer Privacy Protection Act, the Personal Information and Data Protection Tribunal Act, and the Artificial Intelligence and Data Act and to make consequential and related amendments to other Acts"*) (Government of Canada 2022) had its first reading in the House of Commons in 2022. Parts of the bill, such as Part 3, add common requirements for the design, development, and use of artificial intelligence systems, including measures to mitigate the risks of harm and biased output. However, on 6 January 2025, the bill died on the Order Paper, since parliament was prorogued. (Wagner et al. 2024)

In addition to regulatory processes, researchers have also raised questions about the legal duty to find and discover less discriminatory algorithms (Black et al. 2024). Model multiplicity is a phenomenon in which multiple models can achieve equal accuracy on the same task, but differ in their individual predictions or aggregate properties (Black et al. 2022). This means it is possible to have a model with lower disparate impact without affecting the performance. (Black et al. 2024) in their work argues that service providers should have a legal duty to search and discover these LDAs that show less disparate impact with reasonable efforts. There are no known regulations around the discovery of LDAs, and the question of who the burden of proof falls upon is also raised in the paper.

## 5.4 | Generative AI Ownership

For a long time, creativity has been considered one of the most distinguishing attributes of humans. But with the advances in computer science and machine learning, people have tried to add creativity to the machine learning models, and it is where Generative AI appears. Regarding probabilistic modeling, generative models are machine learning models that describe generating a dataset. Sampling from such a model generates new data belonging to the desired distribution (Foster 2022).
**Who is the author?** While current samples of generative models can create high-quality products like meaningful text (OpenAI 2023), different types of art, including paintings (Photoleap 2023), music (Inc. 2023), and executable computer codes, the first generative models which caught the news attention were StyleGAN (Karras et al. 2019) which created hyper-realistic images from human faces and GPT-2 (Radford et al. 2019) that could finish a paragraph of meaningful text related to its received opening sentence.
While the number of generative AI users is increasing daily, a critical question about generative AI ownership still needs to be answered. Consider a situation in which an employee who is expected to add some lines of code to their company's product uses generative AI to generate the code and add it to their product. The question that arises here is who owns the product? Does the company lose copyright over its product because a machine learning model generates one part? What about art generated by generative AI? While an artist who won Colorado State Fair's art competition using AI-generated picture (Harwell 2022) claimed he had not broken any rules, another artist rejected the prize he won at the Sony World Photography award, revealing that AI had generated the winning photo (Grierson 2023). As a prevention from such events, the Grammys forbids AI-generated works from participating in their competitions (King 2023). Debates about the ownership of Generative AI products remain unsolved, and people taking each side of this debate have satisfying reasons. While from one point of view, the owners of the AI-generated products are people who made the models generate such output by their prompts and guidance, the other perspective gives the ownership of these products to model owners who created these generative models or to the models themselves (Epstein et al. 2023).
**Rules and Regulations** Although the debate on generative AI ownership has not concluded, generative AI is being used more daily. Thus, it needs some regulation to decide who is responsible for and who owns the products generated by AI. Based on Federal Register guideline (Register 2023b), the work generated by AI, using only one prompt by the human user, is unprotected. At the same time, if it is a product of interaction between the human and machine where a human makes some modification, selection, or arrangement on the AI product, it is protected by copyright law. Regulations in the EU and UK take a different side toward AI-generated product ownership: a product can be subject to copyright if it is "the author's own intellectual creation." According to Copyright Designs and Patents Act 1988 (CDPA) (Register 2023a), since computer programs cannot be considered as the owner of their product, "the person by whom the arrangements necessary for the creation of the work are undertaken" is known as the author of AI products. As can be seen from different viewpoints of regulations toward generative AI ownership, there has yet to be a finalized committed decision. Some consider AI models as an extension of their human users, and the human owns their products; others take model owners as the creators of the model's products, and the last group considers joint ownership between creators and the model owners. Since the legal landscape around AI ownership is still evolving, a long way exists to achieve commitment.
But what are generative AI's economic and social impacts apart from the ownership of AI products? Based on AI ACT, when products like images, voice, art, etc., are generated by AI and published, it should be exposed that it is AI-generated, especially when they resemble existing persons, places, or events (of European Union 2022). Otherwise,

AI-generated products can easily deceive and manipulate people in various political, social, and cultural domains. Malicious use of AI to defame or create popularity for specific individuals or groups is another important concern toward using Generative AI. In this regard, AI Act added an account for general-purpose AI that "may lead to discriminatory outcomes and the exclusion of certain groups." Product manufacturers or fashion designers can utilize AI-generated content to promote their products without hiring specialized individuals. In this regard, another issue that arises is the economic impact of using generative AI. Despite the significant concerns raised about generative AI's impact on the future of jobs by enabling enterprises to use AI instead of their human forces, the economic consequences of its use and the possibility of further economic inequality due to AI utilization exist. Since launching and utilizing generative AI requires high-level hardware capabilities and software knowledge, equal access to it is impossible for all countries and all individuals within a society. This can significantly increase social and economic inequality between developing and developed countries and between organizations and individuals with financial and informational resources and those without them. The lack of consistent regulation for generative AI added to the concerns about the harms unregulated AI can bring to society (Bender et al. 2021) has resulted in an open letter (for Life 2023b) initiated by Future for Life and signed by more than 3000 people in the technology industry, asking for a six-month pause on giant AI experiments by all companies until sufficient policies are made and enforced. On the other hand, since the AI act is forcing more strict regulations for using generative AI, some European companies claim that it may "jeopardize technological sovereignty" and signed an open letter (for Life 2023a) asking the EU to reconsider its plans to let European companies participate in the advancement of AI.

# 6 | ENVIRONMENTAL IMPACT

While machine learning algorithms are not inherently bad for the climate and environment, their implementation can have certain implications that need to be considered (Zhong et al. 2023b, Patterson et al. 2021, Schwartz et al. 2020). One significant factor is the high energy consumption associated with training and running machine learning models, particularly when GPUs (Graphics Processing Units) are used for accelerated computation. From an environmental viewpoint, the following aspects have to be considered.

**Energy Consumption and Carbon Emissions.** Machine learning algorithms, particularly deep learning models, require extensive computational resources for training (Patterson et al. 2021). This process involves performing numerous complex calculations on large datasets. GPUs are commonly used for their parallel processing capabilities, which accelerate the training process. However, GPUs are power-hungry devices that consume substantial amounts of electricity. The energy consumption of AI algorithms is a significant concern due to the scale at which these algorithms are deployed. The study by Strubell et al. (2019) discovered that the greenhouse gas emissions produced from some of the NLP algorithms were comparable to those from 300 flights traveling between New York and San Francisco. Training state-of-the-art models can take weeks or even months, consuming a substantial amount of energy during that time. The energy consumption is directly proportional to the size and complexity of the model and the amount of data being processed. As machine learning applications become more prevalent across industries, the collective energy consumption associated with training and running increases (Sarswatula et al. 2022).

**Carbon Emissions.** Machine learning algorithms' energy consumption directly impacts carbon emissions (Anthony et al. 2020, Lacoste et al. 2019, Henderson et al. 2020)., as most electricity worldwide still comes from non-renewable sources like coal, natural gas, and oil. These fossil fuels emit CO2 and other greenhouse gases when burned for energy, contributing to climate change. Machine learning algorithms, which heavily use GPUs, often draw electricity from grids powered by fossil fuel-based plants. This results in significant carbon emissions from both the direct power consumption and the indirect emissions from fossil fuel extraction, production, and transportation needed to meet energy demands

**Electronic Waste and Disposal Challenges.** The rapid advancement of machine learning technology necessitates frequent hardware upgrades, including GPUs (Gao et al. 2020). As newer and more powerful GPUs are introduced to the market, older models become obsolete. This cycle of hardware replacement results in electronic waste generation. The electronic waste consists of discarded electronic devices, including GPUs, that are no longer in use. Improper handling and disposal of e-waste pose significant environmental risks. Electronic devices contain hazardous materials, such as lead, mercury, cadmium, and brominated flame retardants, which can contaminate the environment if not managed properly. E-waste often ends up in landfills, where toxic substances can leach into the soil and water, posing threats to ecosystems and human health. Managing e-waste from machine learning infrastructure can be particularly challenging due to the rapid obsolescence of hardware and the need for specialized recycling processes to handle complex electronic components (Miller 2022).

## 6.1 | Using AI for Mitigating Climate Change

Contrary to the potential harms of climate change due to AI, recent research (Cowls et al. 2021, Leal Filho et al. 2022, Coeckelbergh 2021, Huntingford et al. 2019) has gained interest in harnessing AI to combat climate change. By leveraging AI's capabilities, climate modeling and prediction, optimization of resource management and energy efficiency, advancement of renewable energy solutions, and enhancement of climate change adaptation and resilience can be improved. The potential of AI to address the pressing issues of climate change offers a promising path toward sustainability and a greener future.

**Improved Climate Modeling and Prediction.** AI has revolutionized climate modeling by providing powerful tools to process vast amounts of data and make accurate predictions (Schneider et al. 2023, Barnes et al. 2019). Machine learning algorithms can analyze historical climate data, satellite imagery, and environmental variables, unveiling complex climate system dynamics. This enhanced understanding enables us to anticipate extreme weather events, assess the impacts of human activities on the environment, and predict long-term climate trends. AI-driven climate modeling empowers policymakers (Coeckelbergh 2021) with valuable insights to develop effective mitigation and adaptation strategies, fostering a more sustainable future.

**Optimized Resource Management and Energy Efficiency.** AI has the potential to transform resource management and enhance energy efficiency across various sectors (Ahmed et al. 2022, Zeng et al. 2021). By employing AI, especially RL algorithms, smart grids can analyze real-time data on energy consumption, demand, and weather conditions (Ali and Choi 2020, Mazhar et al. 2023). This analysis facilitates efficient energy distribution, minimizes wastage, and encourages the integration of renewable energy sources. Machine learning techniques can optimize energy-intensive processes in industries, transportation systems, and buildings, resulting in significant energy savings.

**Advancements in Renewable Energy Solutions.** Renewable energy plays a pivotal role in mitigating climate change, and AI can accelerate its adoption. AI algorithms can optimize the integration and operation of renewable energy (Chen et al. 2021, Ramachandran 2020, Al-Othman et al. 2022) systems like solar, wind, and hydroelectric power. By analyzing geographical and climatic data, AI can identify optimal locations for solar panels and wind turbines, maximizing energy output. Additionally, AI can aid in developing advanced materials and technologies for efficient energy storage, addressing one of the primary challenges in renewable energy adoption (Shin et al. 2021). By harnessing AI, the shift to a renewable energy-powered world reduces dependence on fossil fuels and mitigates climate change.

In summary, to ensure the responsible use of machine learning algorithms, stakeholders must consider the environmental implications and strive to adopt sustainable practices. This can include optimizing algorithms for efficiency, using energy-efficient hardware, considering renewable energy sources for computing infrastructure, and promoting responsible e-waste management. By balancing technological advancements with environmental considerations, we can potentially harness the benefits of machine learning while minimizing its potential negative impacts on the climate and environment.

# 7 | CONCLUSION

Over the past few years, AI ethics has reaped significant attention as the ramifications of AI decisions permeate every layer of society. We provide a review to understand its intricacies and challenges through notions of privacy and data protection, transparency and explainability, fairness and equity, responsibility, accountability, and regulations and lastly the environmental impacts. The notions presented in this paper are key ingredients to building trustworthy AI systems. Nevertheless, there is not yet a checklist that, when fulfilled, ensures trustworthy AI. The integration of these principles depends on the problems and the business needs. Moreover, certain ethical principles can be conflicting or open to compromise. For example, prioritizing privacy in a system might lead to diminished performance and fairness (Bagdasaryan et al. 2019, Uniyal et al. 2021), introducing trade-offs among accuracy, fairness, and privacy. Determining which ethical aspect should take precedence in such scenarios remains ambiguous. Regulators are expected to establish a baseline for fairness and privacy, necessitating compliance audits before AI system deployment (Falco et al. 2021).

Concurrently, constructing fair models introduces privacy concerns related to demographic information (Ferry et al. 2023, Lyu and Chen 2021), or faces challenges when demographic information is unavailable due to privacy constraints (Hashimoto et al. 2018, Kenfack et al. 2023a). While explainability is crucial for instilling trust, disparities in the quality of explanations (Dai et al. 2022) and the risk of 'washing' (Aïvodji et al. 2019) persist.

Assessments of advancements in building models robust to attacks predominantly focus on accuracy, though other principles such as fairness are also vulnerable to attacks (Mehrabi et al. 2021).

Conversely, promoting accountability in companies, with clear consequences for individuals involved in the AI lifecycle when unexpected outcomes arise, could instill fear of professional repercussions and potentially stifle innovation. Therefore, it is imperative to formulate policies that strike a balance between fostering innovation and ensuring accountability (Millar et al. 2018, Ammanath 2022). Despite these challenges requiring considerable attention, an expanding body of work is dedicated to addressing them (Papernot et al. 2021, Tran et al. 2022, Li et al. 2023b, Wang et al. 2021, Xu et al. 2021).

Its interdisciplinary scope and emphasis on actionable insights make it a valuable resource for researchers, developers, and policymakers alike. Still, several open challenges remain, such as reconciling ethical trade-offs, operationalizing fairness without compromising privacy, and developing globally inclusive governance frameworks. Future research should focus on creating standardized evaluation metrics, scalable mitigation techniques, and participatory design processes to ensure responsible and reproducible AI across diverse contexts.

To sum up, while this paper investigates the interdisciplinary scope of AI Ethics from classical machine learning to foundation models, it highlights open challenges, including ethical trade-offs, the need for a global, inclusive regulatory and governance framework, and the environmental impact of AI.

## SUPPORTING INFORMATION

Additional supporting information may be found in the online version of the article at the publisher's website.